\documentclass{jfm}

\usepackage{graphicx}
\usepackage{natbib}
\usepackage{color}
\usepackage{amsfonts,amssymb,amsmath}
\usepackage{bm}

\graphicspath{{./}}

\ifCUPmtlplainloaded \else
  \checkfont{eurm10}
  \iffontfound
    \IfFileExists{upmath.sty}
      {\typeout{^^JFound AMS Euler Roman fonts on the system,
                   using the 'upmath' package.^^J}%
       \usepackage{upmath}}
      {\typeout{^^JFound AMS Euler Roman fonts on the system, but you
                   dont seem to have the}%
       \typeout{'upmath' package installed. JFM.cls can take advantage
                 of these fonts,^^Jif you use 'upmath' package.^^J}%
      }
  \else
  \fi
\fi

\ifCUPmtlplainloaded \else
  \checkfont{msam10}
  \iffontfound
    \IfFileExists{amssymb.sty}
      {\typeout{^^JFound AMS Symbol fonts on the system, using the
                'amssymb' package.^^J}%
       \usepackage{amssymb}%

      }{}
  \fi
\fi

\ifCUPmtlplainloaded \else
  \IfFileExists{amsbsy.sty}
    {\typeout{^^JFound the 'amsbsy' package on the system, using it.^^J}%
     \usepackage{amsbsy}}
    {}
\fi

\newsavebox{\astrutbox}
\sbox{\astrutbox}{\rule[-5pt]{0pt}{20pt}}

\newcommand{\BS}[1]{{\color{black}#1}}

\title[turbulent band in plane Poiseuille flow]{The growth mechanism of turbulent bands in channel flow at low Reynolds numbers}

\author[Xiangkai Xiao, Baofang Song]
{Xiangkai Xiao, Baofang Song
\thanks{Email address for correspondence: baofang\_song@tju.edu.cn}}
\affiliation{Center for Applied Mathematics, Tianjin University, Tianjin 300072, China %\\[\affilskip]
}

\date{?; revised ?; accepted ?. - To be entered by editorial office}
\begin{document}

\maketitle

\begin{abstract}
In this work, we carried out direct numerical simulations in large channel domains and studied the kinematics and dynamics of \BS{fully localised} turbulent bands at Reynolds number $Re=750$. Our results show that the downstream end of the band features fast streak generation and travels into the adjacent laminar flow, whereas streaks at the upstream end continually and relatively more slowly decay. This asymmetry is responsible for the transverse growth of the band. We particularly investigated the mechanism of streak generation at the downstream end, which engines the growth of the band. We identified a spanwise inflectional instability associated with the local mean flow near the downstream end and our results strongly suggest that this instability is responsible for the streak generation and ultimately for the growth of the band. Based on our study, we propose a possible self-sustaining mechanism of \BS{fully localised} turbulent bands at low Reynolds numbers in channel flow.
\end{abstract}


\section{Introduction}
The laminar state of channel flow becomes linearly unstable above Re=5772, however, given sufficiently strong perturbations, the flow can undergo subcritical transition to turbulence and become sustained at much lower Reynolds number at $Re\simeq 660$ \citep{Tao2018,Xiong2015}. Here, the Reynolds number is defined as $Re=U_ch/\nu$, where $U_c$ is the centerline velocity of the unperturbed parabolic flow, $h$ the half gap-width and $\nu$ the kinematic viscosity of the fluid. At low Reynolds numbers, \citet{Tsukahara2005} first found that channel flow turbulence appears as localised bands tilted with respect to the streamwise direction, and many studies have investigated band structures in channel flow since then \citep{Tsukahara2014a,Tsukahara2014b,Tuckerman2014,Xiong2015,Tao2018,Kanazawa2018}. Similar band patterns also appear in the transitional regime in other plane shear flows \citep{Coles1965,Prigent2002,Barkley2005,Duguet2010,Duguet2013,Chantry2017}. 

The transition scenario through localised turbulence in channel flow has attracted much attention in recent studies. \citet{Sano2016} proposed a directed percolation (DP) transition, above $Re_{\mathrm {cr,DP}}\simeq 830$, to featureless turbulence at much larger Reynolds numbers. Together with \citet{Lemoult2016, Chantry2017, Mukund2018}, recent studies categorised the transition to sustained turbulence in shear flows into the DP universality class. DP theory predicts no sustained turbulence below the DP critical point. However, using large computational domains, \citet{Xiong2015,Tao2018} recently reported that, even below $Re=830$, a single turbulent band can persistently grow transversely in the absence of interactions with other bands and channel side walls. Therefore, the turbulence fraction can be arbitrarily small if a single band persists in an arbitrarily large domain. This sparse-turbulence state was argued to be responsible for the deviation of the transition scenario in channel flow from the DP theory of \citet{Sano2016}. \citet{Shimizu2018} studied interactions between multiple bands in a very large domain ($500h\times 2h\times 250h$) and also reported sustained banded turbulence structures below the DP critical point of \citet{Sano2016}. Further, through observing and modeling of the complex interactions between bands, they proposed a bifurcation process that leads the one-sided banded turbulence state at lowest $Re$ to two-sided banded turbulence as $Re$ increases, and eventually to the onset of DP at $Re=976$, a bit higher than the DP critical point of \citet{Sano2016}. 

However, there have been so far rather limited studies of the detailed flow structures and the dynamics of individual turbulent bands in channel flow, especially in large flow domains. The mean flow of the repeated turbulent band-laminar gap pattern has been studied and modeled by \citet{Tuckerman2014}. Regarding the structure and dynamics of a fully localised band in a large domain, \citet{Shimizu2018, Kanazawa2018} noticed that a turbulent band at low Reynolds numbers is characterized by an active head at the downstream end and a rather diffusive tail at the upstream end. 
The numerical simulation of \citet{Shimizu2018} in a $500h\times 2h\times 250h$ domain up to $\mathcal{O}(10^5)$ time units seems to suggest that a single band does not grow infinitely long, rather can undergo longitudinal splitting, which seems to disagree with the conclusion of \citet{Tao2018}. Nevertheless, their results both indicated that a single band can be sustained at low Reynolds numbers.
To our best knowledge, there have been no studies regarding the growth mechanism and the self-sustaining mechanism of an individual turbulent band in channel flow. In this contribution, we aim to fill this gap and investigate the kinematics and particularly the growth mechanism of fully localised bands in more detail at $Re=750$.

\section{Methods}
We solve the non-dimensional incompressible Navier-Stokes equations
\begin{equation}\label{equ:NS}
 \frac{\partial \bm u}{\partial t}+{\bm u}\cdot\bm{\nabla}
{\bm u}=-{\bm{\nabla}p}+\frac{1}{Re}{\bm\nabla^2}{\bm u}, \;
\bm{\nabla}\cdot{\bm u}=0
\end{equation}
for a volume flux-driven channel flow in Cartisian coordinates $(x, y, z)$, where $\bm u$ denotes velocity, $p$ denotes pressure and $x$, $y$ and $z$ represent the streamwise, wall-normal and spanwise coordinates, respectively. \BS{The flux in the streamwise direction is fixed to be that of the unperturbed parabolic laminar flow and the flux in the spanwise is zero.} Velocities are normalised by $U_c$, length is normalized by $h$ and time by $h/U_c$. The origin of the y-axis is placed at the channel center. No-slip boundary condition, \BS{$\bm u=0$}, for velocities is imposed at channel walls\BS{, i.e., at $y=\pm 1$}.
Periodic boundary conditions are imposed in streamwise and spanwise directions. A hybrid Fourier spectral-finite difference method is used to solve Eqs. (\ref{equ:NS}). 
In the wall-normal direction, a finite-difference method with a 9-point stencil is employed for the discretisation.
Therefore, the velocity and pressure fields can be expressed as
\begin{equation}
{A}(x,y,z,t)=\sum_{k=-K}^{K}\sum_{m=-M}^{M}\hat{A}_{k,m}(y,t)e^{i(\alpha kx+\beta mz)},
\end{equation}
where $k$ and $m$ are the indices of the streamwise and spanwise Fourier modes, respectively, $\hat A_{k,m}$ is the Fourier coefficient of the mode $(k,m)$ and $\alpha$ and $\beta$ are the fundamental wave numbers in the streamwise and spanwise directions, respectively. The size of the computational domain is set to be $L_x=2\pi/\alpha$ and $L_z=2\pi/\beta$. We use the finite difference scheme and the parallelisation strategy of openpipeflow \citep{Willis2017} and adopt the time-stepping and projection scheme of \citet{Hugues1998} for integrating the incompressible system.

\section{Results}
We first performed direct numerical simulations (DNS) at $Re=750$ in a computational box with $L_x=120$ and $L_z=120$. We used 768 Fourier modes ($K=M=384$) in both streamwise and spanwise directions. \BS{This resolution gives 6 grid points over $h$ and is comparable with \citet{Tao2018} and results in a decrease by more than 4 orders of magnitude in the amplitude of the Fourier coefficient from the lowest to the highest Fourier mode for a turbulent field. \citet{Tuckerman2014} used 12 grid points over $h$ for $Re$ up to 2300, which also implies that our resolution is sufficient for $Re=750$ assuming $N\sim Re^{3/4}$, where $N$ is the number of grid points in one spatial dimension.} We used 72 Chebyshev grid points for the finite difference discretisation in the wall normal direction (we also tested 96 points and found 72 sufficient). A time-step size of $\Delta t=0.01$ is used for the time integration, which is sufficiently small for this low Reynolds number.

\begin{figure}
\centering
\includegraphics[width=0.7\textwidth]{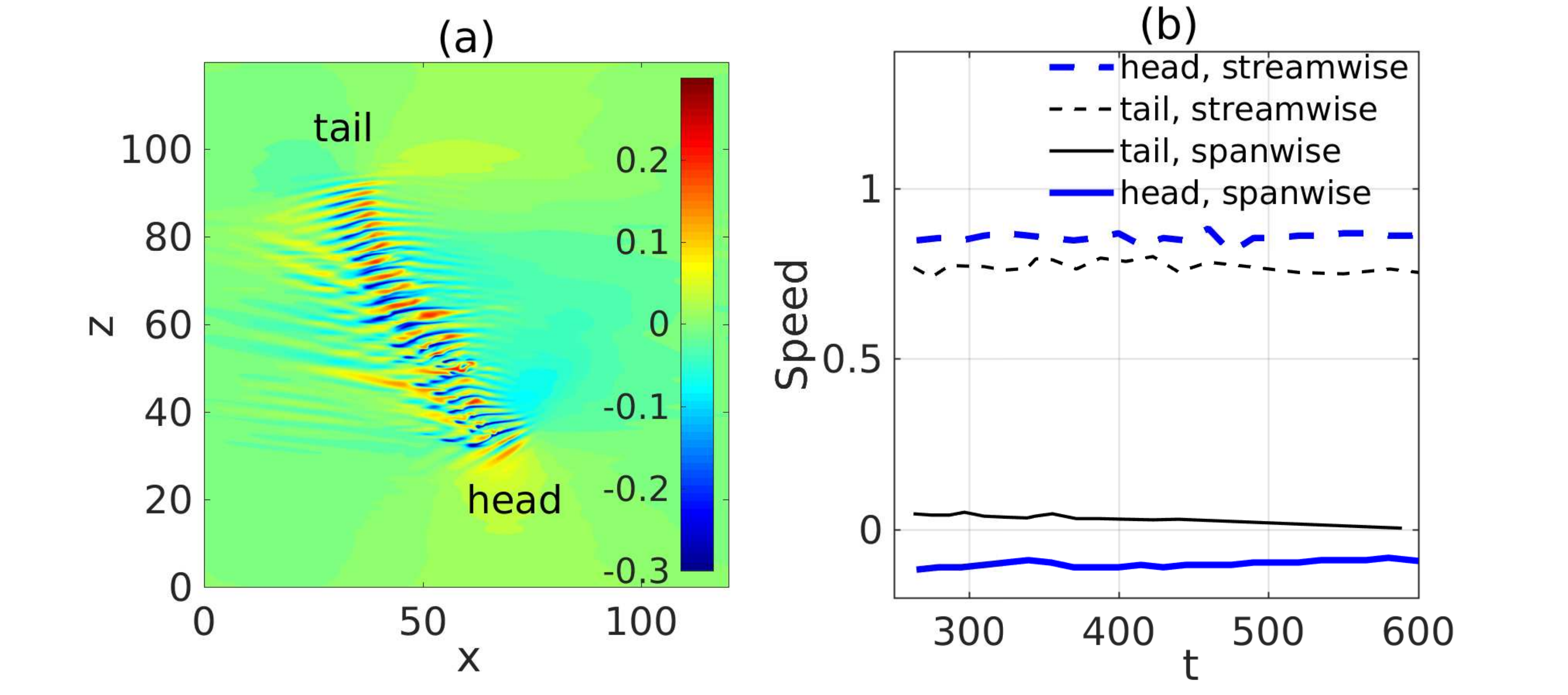}
%\vspace*{8pt}
\caption{\label{fig:band_and_head_tail_speed} (a) The turbulent band formed at Re=750 starting from an initial condition simulated at $Re=950$. The streamwise velocity $u_x$ in the $x$-$z$ cut-plane at $y=-0.5$ is plotted as the colormap. \BS{The flow is from left to right in the figure and the band is tilted about the streamwise direction at an angle of 57 degrees at this time instant, which is calculated using the method given in \citet{Tao2018}.} (b) The streamwise and spanwise speeds of the tail and head of the band shown in (a).}
\end{figure}

We first perturbed the flow at $Re=950$ using a localized vortical perturbation proposed by \citet{Henningson1991}. After a band was formed, we reduced the Reynolds number to $Re$=750 and, after some initial adjustment, obtained a single band, see Fig. \ref{fig:band_and_head_tail_speed}(a).  
As mentioned by \citet{Shimizu2018}, the turbulent band in this Reynolds number regime is characterised by an active downstream end (referred to as {\it head} hereafter) and a rather diffusive upstream end (referred to as {\it tail} hereafter), which are labeled in the figure. In addition, we observed a wave-like pattern, i.e. a pattern of alternating high and low speed streaks, localised in the streamwise direction, along the turbulent band in the $x$-$z$ cut-plane at $y=-0.5$ (see Fig. \ref{fig:band_and_head_tail_speed}(a)). \BS{Similar wave-pattern was also observed at the downstream end of turbulent bands in the numerical simulations of \citet{Kanazawa2018}, at the wingtip of turbulent spots at higher Reynolds numbers in channel flow \citep{Carlson1982,Alavyoon1986,Henningson1987,Henningson1989,Henningson1991} and in plane Couette flow \citep{Dauchot1995} as well.}  
\subsection{Kinematics of the head and tail} 

To understand the growth mechanism of the band, we first investigated the kinematics of the head and tail. We tracked the head and tail and measured their mean speeds, which are shown in Fig. \ref{fig:band_and_head_tail_speed} (b). \BS{Streamwise and spanwise positions of the head and tail are determined by setting a proper threshold in the fluctuation intensity above which the region is considered turbulent and otherwise laminar. Speed is subsequently calculated based on the time series of the position.} Both streamwise and spanwise speeds of the head are quite stable over the time window we investigated, and the former is 0.85 and the latter is -0.10 on average (see the bold lines).
Clearly the head is invading the adjacent laminar flow. The streamwise speed of the tail is also quite stable with an average of 0.76 (see the dashed thin line), whereas the spanwise speed of the tail is about 0.035 at $t$=250, and slowly decreases to nearly zero at about $t$=600 (see the solid thin line). Clearly, the spanwise speed of the tail has not stabilised within the time window of about 300 time units and shows a decreasing trend as time goes on. The domain size chosen is not large enough for a longer measurement time because at later times the head and tail get too close to each other and start to interact due to the periodic boundary condition. 

\begin{figure}
\centering
\includegraphics[width=0.8\textwidth]{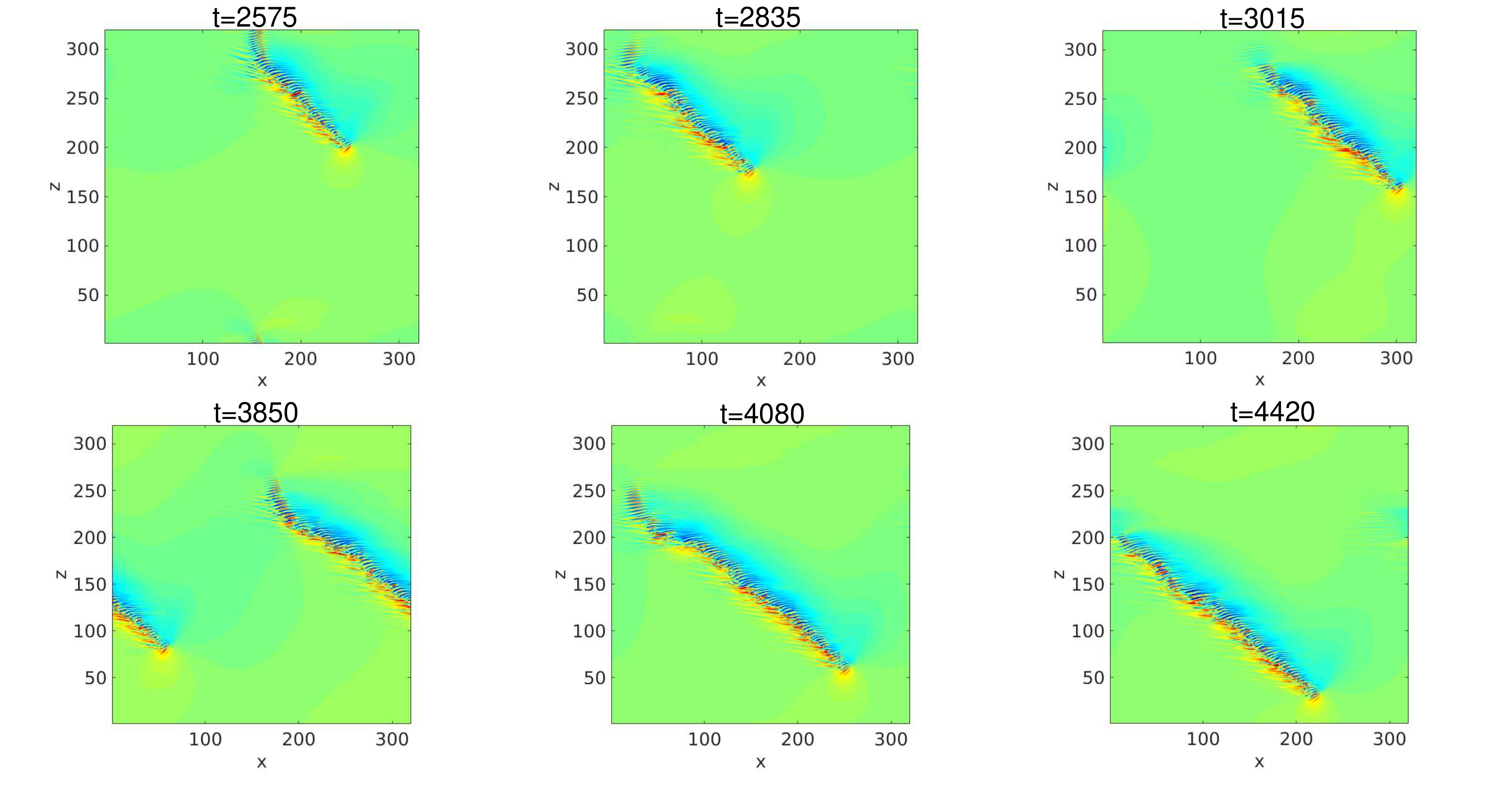}
%\vspace*{8pt}
\caption{The turbulent band in the large domain with $L_x=L_z=320$. The contours of streamwise velocity are visualized in the $x$-$z$ cut-plane at $y=-0.5$. Flow is from left to right. \label{fig:tail_large_domain}}
\end{figure}

To study the kinematics of the tail with a less restrictive domain size, we considered a larger computation domain with $L_x=L_z=320$. For this domain size, we used a resolution of 2048 Fourier modes ($K=M=1024$) in both spanwise and streamwise directions \BS{($h/\Delta x=h/\Delta z\simeq 6.4$)}. Figure \ref{fig:tail_large_domain} shows the development of a turbulent band in the domain. This large domain allows a much longer observation time of the band without significant head-tail interaction due to the periodic boundary condition. 

Overall, the tail is moving downward in the spanwise direction, i.e., the spanwise speed is negative, which cannot be observed in the small domain shown before. Nevertheless, this negative speed is consistent with the trend of the spanwise speed shown in Fig. \ref{fig:band_and_head_tail_speed}(b). However, because the tail continually decays as a bulk (multiple streaky structures decay as a whole, see the tail shown in panels from $t=2575$ to 3015 and from $t=3850$ to 4420 in Fig. \ref{fig:tail_large_domain}.), it is difficult to define a characteristic advection speed. Nevertheless, based on the spanwise position of the tail over very large time interval ($\mathcal{O}(10^3)$), we can estimate an average spanwise speed of the tail, which is roughly -0.06. 

\begin{figure}
\centering
\includegraphics[width=0.85\textwidth]{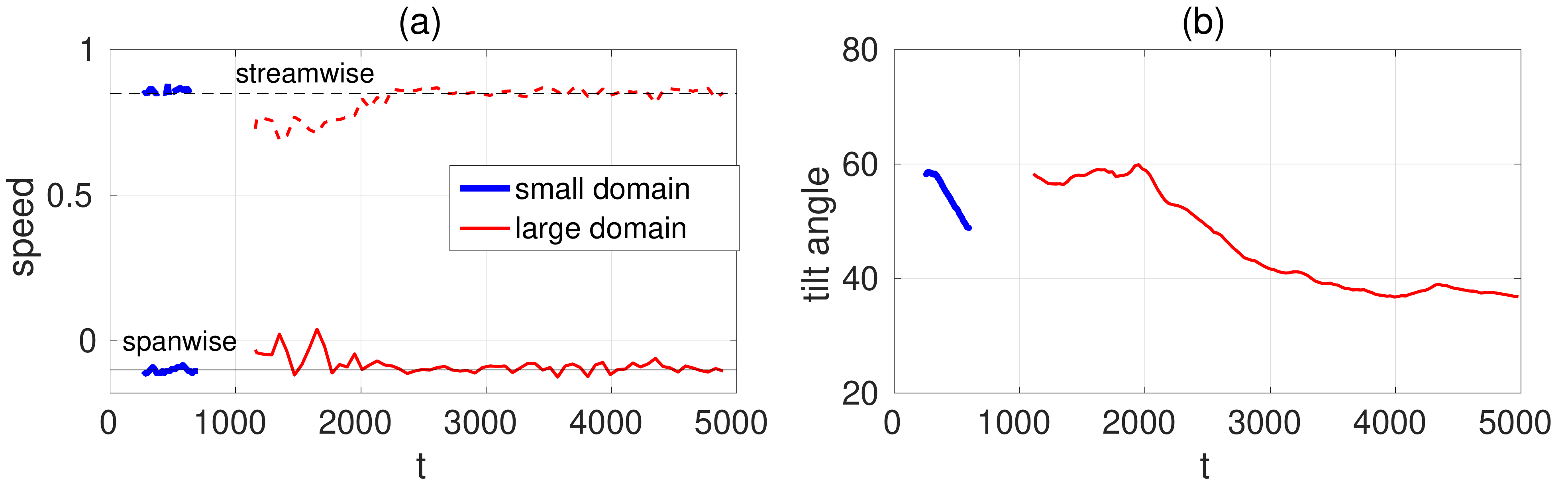}
%\vspace*{8pt}
\caption{\label{fig:speed_tilt_angle_small_large}\BS{The speeds of the head (a) and the tilt angle of the turbulent band (b) for the small domain ($120\times 2\times 120$) and large domain ($320\times 2\times 320$). The reference speeds of 0.85 and -0.1 are plotted as thin dashed and solid lines. The tilt angle shown in panel (b) is calculated using the method used by \citet{Tao2018}.}}
\end{figure}

It is noted that the kinematics of the head is not affected by the domain size, as both the streamwise and spanwise speed of the head are still 0.85 and -0.1 on average, respectively, \BS{see the $t>2000$ part of the large domain case in Fig. \ref{fig:speed_tilt_angle_small_large}(a)}. This implies that the head has rather robust structure and dynamics. 
\BS{The tilt angle of the band is also calculated for the two domain sizes using the method proposed by \citet{Tao2018} and compared in Fig. \ref{fig:speed_tilt_angle_small_large}(b). It can be seen that the tilt angle keeps decreasing and has not stabilised in the whole observation time in the small domain. In the large domain, we can see that the tilt angle seems to have stabilised and fluctuate between 37 and 39 degrees after a very long transient of about 2000 time units (the length of the transient is probably initial condition dependent as \citet{Tao2018} showed a shorter transient of about 1000 time units).}

\subsection{Streak generation at the head of the band}

\begin{figure}
\centering
\includegraphics[width=0.85\textwidth]{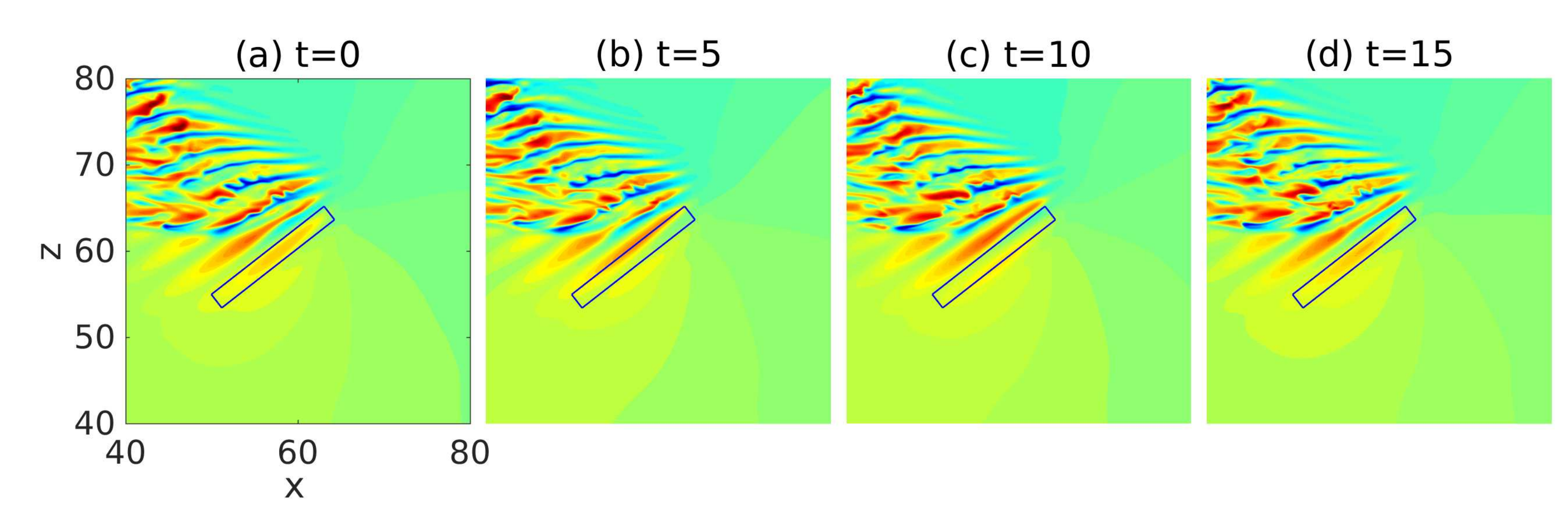}
%\vspace*{8pt}
\caption{The illustration of the generation of streaks at the band head. A black rectangle is used to mark the first observable streak generated at the head. Streamwise velocity fluctuations are plotted as the colormap. Panel (a) to (d) show four time instants separated by 5 time units. \label{fig:streak_generation_first_streak}}
\end{figure}
To investigate the dynamics of the head, we went into the frame of reference co-moving with the head at a spanwise speed of -0.1 and visualised the flow in the $x$-$z$ cut-plane at $y=-0.5$, see Fig. \ref{fig:streak_generation_first_streak}. We observed that wave-like structures, i.e., alternating low and high speed streaks, are continually generated at the head. The high speed streak in the black rectangle (panel (a)) moves toward the band body and leaves the rectangle (panel (b, c)), and subsequently a new high speed streak forms at the head and appears in the rectangle after roughly 15 time units (panel (d)). Noting that the enclosed streak in (d) is nearly the same as the one in (a), we estimated that the period of the streak generation is approximately 15 time units. See Online Movie 1 for a better visualisation of this dynamical process. \BS{\citet{Kanazawa2018} also observed this process in channel flow and attributed the growth of a turbulent band to the periodic wave generation at the head.}

\subsubsection{Mean flow around the head}
\begin{figure}
\centering
\includegraphics[width=0.95\textwidth]{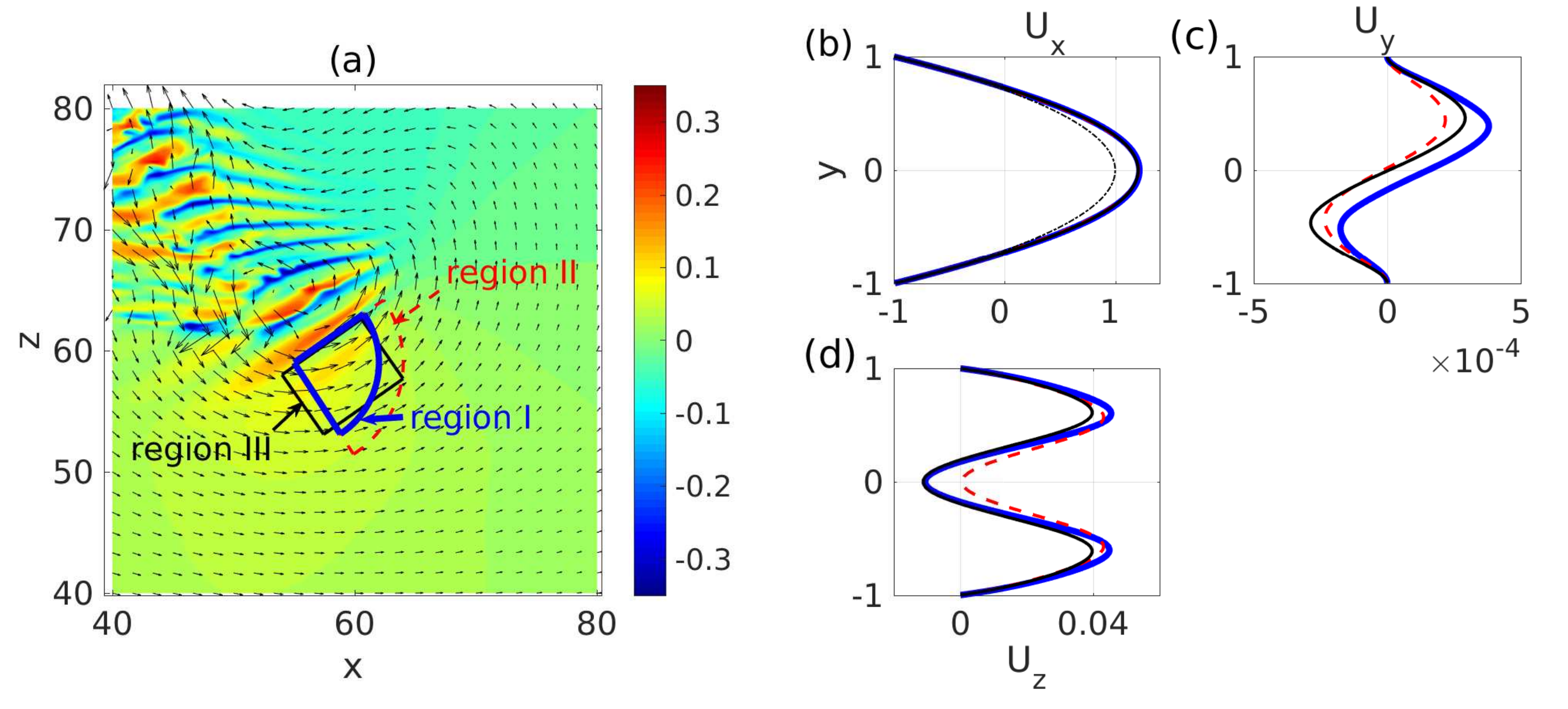}
%\vspace*{8pt}
\caption{(a) Three regions near the head in which the spatially and temporally averaged velocity profiles are calculated: region I (bold blue), II (dashed red) and III (thin black). (b-d) The averaged velocity profiles of $U_x$, $U_y$ and $U_z$ in the three regions as defined in panel (a). The color and line style correspond to those in (a). \BS{In panel (b), the parabola is shown as a black thin dash-dotted line.} \label{fig:domain_decomposition_mean_profiles}}
\end{figure}

\BS{For turbulent spots at higher Reynolds numbers, \citet{Li1989,Henningson1987,Henningson1989} proposed that the generation of wave-like structures at the wingtip of the spot is due to the destabilisation of the flow by the spot in the wingtip region.  
Following their studies, 
we investigated the temporally and spatially averaged mean flow at the head (referred to as `mean flow' hereafter)}, particularly in the region where the first visible streak is periodically generated. We selected three regions enclosing the first visible high speed streak, and named them as region I (bold blue), region II (dashed red) and region III (thin black), see Fig. \ref{fig:domain_decomposition_mean_profiles}(a). We considered different shapes (quarter circle and rectangle) and different positions and sizes of the region, in order to check the sensitivity of the mean flow to the selected region. We averaged the flow both spatially and temporally in the three regions in the co-moving frame of reference using about 150 snapshots separated by $\Delta t=1$, and obtained the one-dimensional mean flow. The duration of the averaging is one order of magnitude larger than the estimated streak generation period. We plotted the streamwise, wall normal and spanwise velocity profiles of the mean flow in the three regions in Fig. \ref{fig:domain_decomposition_mean_profiles}(b-d). In fact, the instantaneous spatially averaged velocity profile fluctuates slightly due to weak nonlinearity and fluctuations in the position of the head. See Online Movie 2 for the temporal behaviour of the spatially averaged velocity profile in region I.

The mean streamwise velocity profile is considerably higher than the basic parabolic profile, indicating a higher local Reynolds number. The mean spanwise velocity profiles, in all three regions, exhibit strong inflection, \BS{similar to the observation of \citet{Henningson1987,Henningson1989} in the wingtip region of turbulent spots}. The $y$-component of the mean flow velocity is orders of magnitude smaller than the other two. We speculate that the streaky structures at the head and within the band body originate from a local inflectional instability at the head, \BS{as proposed by \citet{Henningson1987,Henningson1989} for spots}. To check this possibility, we carried out a linear stability analysis based on the measured mean flow.

\subsubsection{Linear stability analysis}
\label{sec:linear_analysis}
Considering that the $y$-component of the averaged mean flow is orders of magnitude smaller than the other two (see Fig. \ref{fig:domain_decomposition_mean_profiles}(b-d)), we zeroed the $y$-component and considered a two-component mean flow $\bm U_b(y)=(U_x(y), 0, U_z(y))$, which is a solenoidal field and only depends on $y$. We adopted the velocity-vorticity formulation of the Navier-Stokes equations, and derived the governing equations of small perturbations with respect to $\bm U_b$ as the following:
\begin{eqnarray}
 \left(\frac{\partial}{\partial t}+U_x\frac{\partial}{\partial x} + U_z\frac{\partial}{\partial z}\right)\nabla^2u_y - \frac{\mathrm{d}^2U_x}{\mathrm{d}y^2}\frac{\partial u_y}{\partial x}-\frac{\mathrm{d}^2U_z}{\mathrm{d}y^2}\frac{\partial u_y}{\partial z}& =\mathrm{\frac{1}{Re}}\nabla^4u_y, \\
 \left(\frac{\partial}{\partial t}+U_x\frac{\partial}{\partial x}+ U_z\frac{\partial}{\partial z}\right)\eta + \frac{\mathrm{d}U_x}{\mathrm{d}y}\frac{\partial u_y}{\partial z} -\frac{\mathrm{d}U_z}{\mathrm{d}y}\frac{\partial u_y}{\partial x}  &=\mathrm{\frac{1}{Re}}\nabla^2\eta,
\end{eqnarray}
where $u_y$ denotes the wall normal velocity and $\eta=\partial u_x/\partial z-\partial u_z/\partial x$ is the $y$-component of the fluctuating vorticity \BS{(for an alternative formulation, see \citet{Henningson1987})}. The boundary condition is $u_y=\partial u_y/\partial y=0$ and $\eta=0$ at $y=\pm 1$. A Chebyshev-Fourier-Fourier spectral method is used to discretise this system, for details see \citet{Trefethen2000}.  

\begin{figure}
\centering
\includegraphics[width=0.85\textwidth]{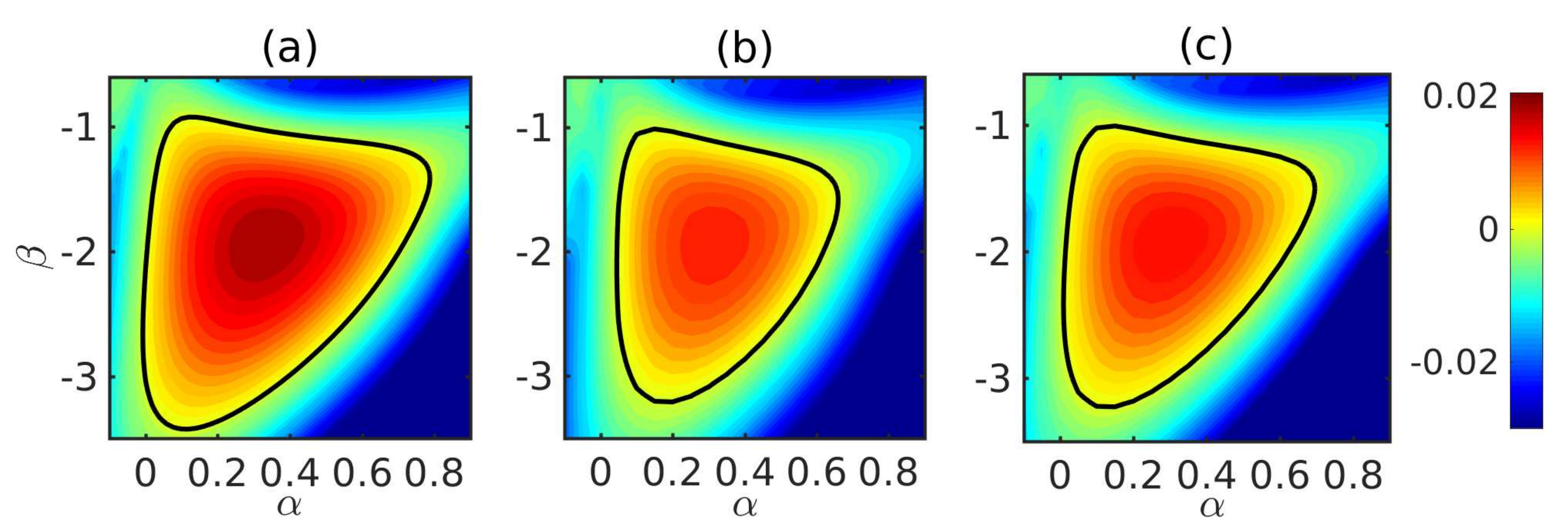}
%\vspace*{8pt}
\caption{The distribution of the imaginary part of the most unstable eigenvalue in the $\alpha$-$\beta$ wavenumber plane. (a), (b) and (c) correspond to the region I, II and III shown in Fig. \ref{fig:domain_decomposition_mean_profiles}(a), respectively. The black line marks the stability boundary. \label{fig:unstable_regions}}
\end{figure}

The linear stability is analysed mode by mode and contours of the most unstable/least stable eigenvalue are plotted in the $\alpha$-$\beta$ wavenumber plane, see Fig. \ref{fig:unstable_regions}. It is clear that a large region in the wavenumber plane is unstable, with positive imaginary part of the most unstable eigenvalue, for all three averaged mean flows (see the region enclosed by the bold black line). 
\begin{figure}
\centering
\includegraphics[width=0.45\textwidth]{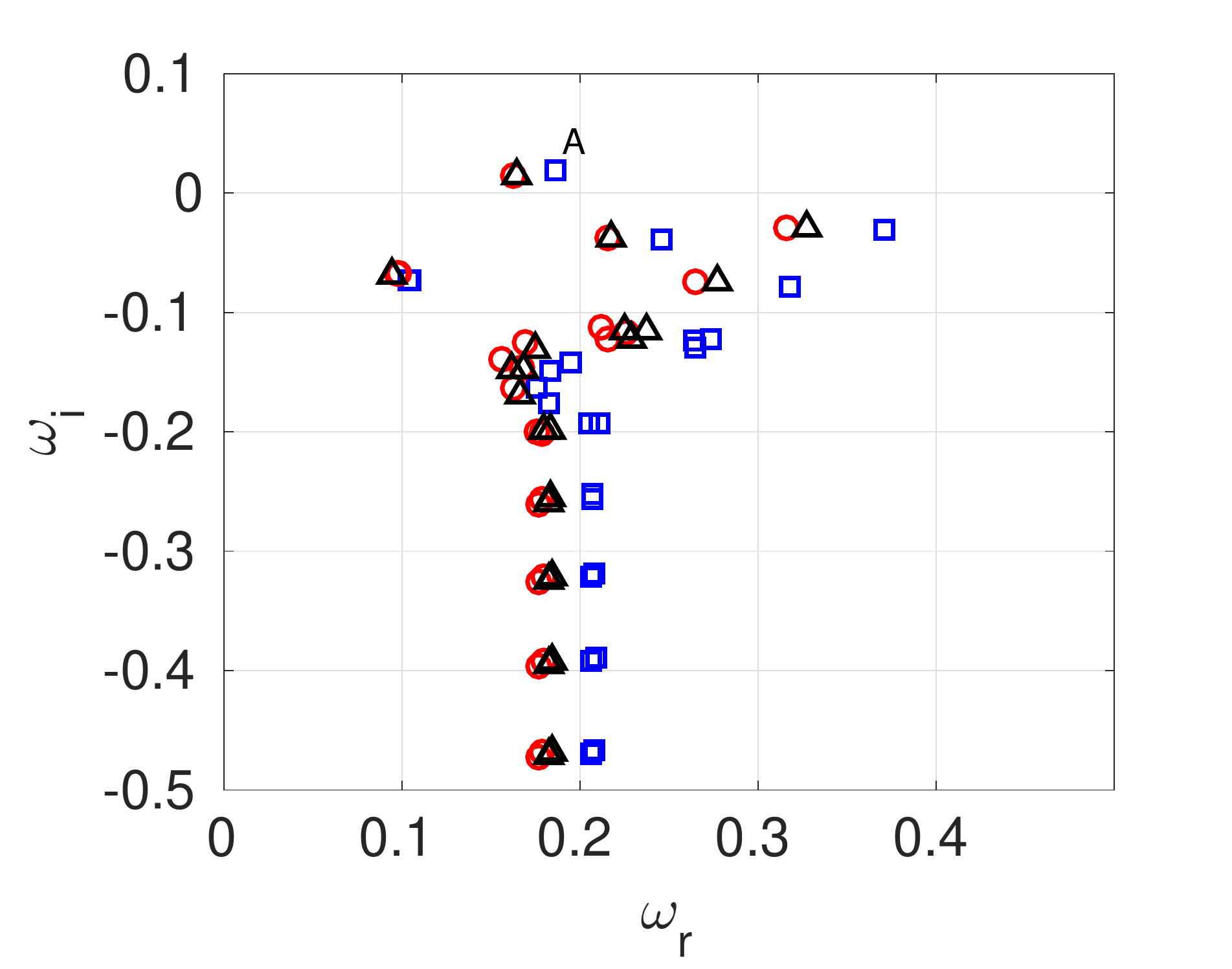}
%\vspace*{8pt}
\caption{The spectrum of the most unstable modes correspond to region I (blue squares), II (red circles) and III (black triangles). The $x$-axis shows the real part $\omega_r$ and the $y$-axis shows the imaginary part $\omega_i$ of the eigenvalues $\omega$. \BS{The flow is linearly unstable if the imaginary part is located in the region of $\omega_i>0$.} The most unstable eigenvalue for region I is labeled by a letter A (the blue square). \label{fig:spectrum}}
\end{figure}
The most unstable mode is $(\alpha,\beta)=(0.34, -1.92)$, $(0.30, -1.94)$ and ($0.30, -1.93$) for region I, II and III, respectively. The most unstable eigenvalues of these three modes are $\omega$=0.1861+0.0192i, 0.1628+0.0140i and 0.1643+0.0148i (see Fig. \ref{fig:spectrum}), respectively, where the real parts give the frequencies and the imaginary parts give the temporal growth rates of the corresponding eigenvectors. From the most unstable eigenvalue and the corresponding wave number, we can calculate the spanwise wave speed of the flow structures given by the associated eigenvector as $\omega_r/\beta$, which is -0.097, -0.084 and -0.085 for region I, II and III, respectively. These speeds are very close to the measured spanwise speed of the head. 

\begin{figure}
\centering
\includegraphics[width=0.6\textwidth]{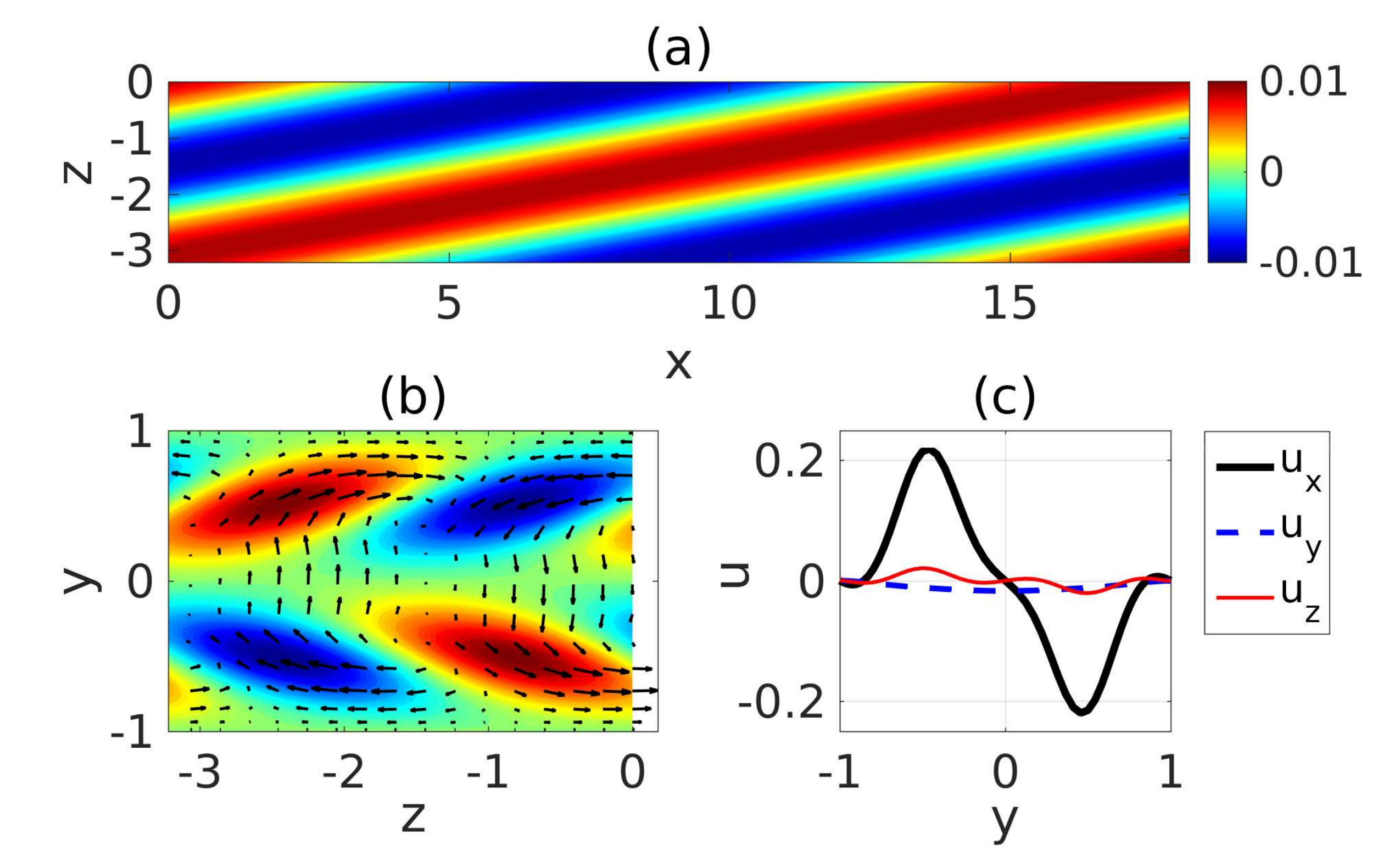}
%\vspace*{8pt}
\caption{The visualisation of the most unstable perturbation corresponding to region I, which is at $(\alpha,\beta)$=(0.34, -1.92). (a) The contours of streamwise velocity component \BS{(with respect to $\bm U_b$)} in the $x$-$z$ cutplane at $y=-0.5$. (b) The contours of streamwise velocity and in-plane velocity vectors shown in the $z$-$y$ cross-section at $x=0$. (c) The velocity profiles of $u_x$ (bold black), $u_y$ (dashed blue) and $u_z$ (thin red) are plotted at $(x,z)$=(0, -1). \label{fig:most_unstable_mode}}
\end{figure}

As shown, the instability properties of the three regions are very close. In the following, we only show analysis of region I. Figure \ref{fig:most_unstable_mode} visualises the eigenvector associated with the most unstable eigenvalue of the most unstable mode for region I (labeled by a letter A in Fig. \ref{fig:spectrum}). Panel (a) shows the streamwise velocity fluctuation \BS{with respect to $\bm U_b$} in the $x$-$z$ cut-plane at $y=0.5$ and (b) shows the same quantity in the $z$-$y$ cross-section at $x=0$. We can observe high and low velocity streak pattern tilted with respect to the streamwise direction. Panel (c) shows the velocity profiles at position $(x, z)$=(0, -1), which indicates a streak-dominant flow with $u_x$ being larger by more than an order of magnitude than other two velocity components.

The most unstable perturbation, as shown in Figure \ref{fig:most_unstable_mode}, has some close similarities with the streaky structures in the real turbulent band.
\begin{enumerate}
\item The instability indeed generates tilted streaky flow pattern. The tilting direction is similar to that of the streaks in the band, see Fig. \ref{fig:band_and_head_tail_speed}(a) and Fig. \ref{fig:streak_generation_first_streak}, although the tilt angle, \BS{which is about 10 degrees}, is smaller \BS{than that of the first two streaks at the head, which is about 38 degrees.}
\item The spanwise wave number of the streaky pattern, 1.92 (absolute value), is very close to that of the streaks at the head of the band, which is \BS{about 2.1} and can be estimated later in Fig. \ref{fig:nonlinear_development}(c).
\item The wave speed in the spanwise direction, -0.097, is very close to the spanwise speed of the head, -0.1.
\end{enumerate}

These results suggest that this spanwise instability is responsible for the generation of streaky structures at the head. 

\subsubsection{Nonlinear development}

To further evidence that the streaky structures are generated by this spanwise inflectional instability, we performed direct numerical simulations of the development of small perturbations given the base flow $\bm U_b$. Here we only investigated region I for the nonlinear analysis. We took the averaged mean velocity profiles in region I as the base flow $\bm U_b$, i.e., we imposed forcing in $x$ and $z$ directions to produce this base flow, and added very small perturbations ($\mathcal{O}(10^{-5})$) on top and integrated the Navier-Stokes equations. 

The base flow will be nearly fixed during the linear development of the perturbations. The initial perturbation was obtained by taking a fully turbulent flow field simulated at a higher Reynolds number, $Re=1500$, and scaling down the velocity fluctuations about the basic parabolic flow by a small factor of $10^{-4}$. In this way, we not only perturbed the most unstable mode but also perturbed other modes.

\begin{figure}
\centering
\includegraphics[width=0.45\textwidth]{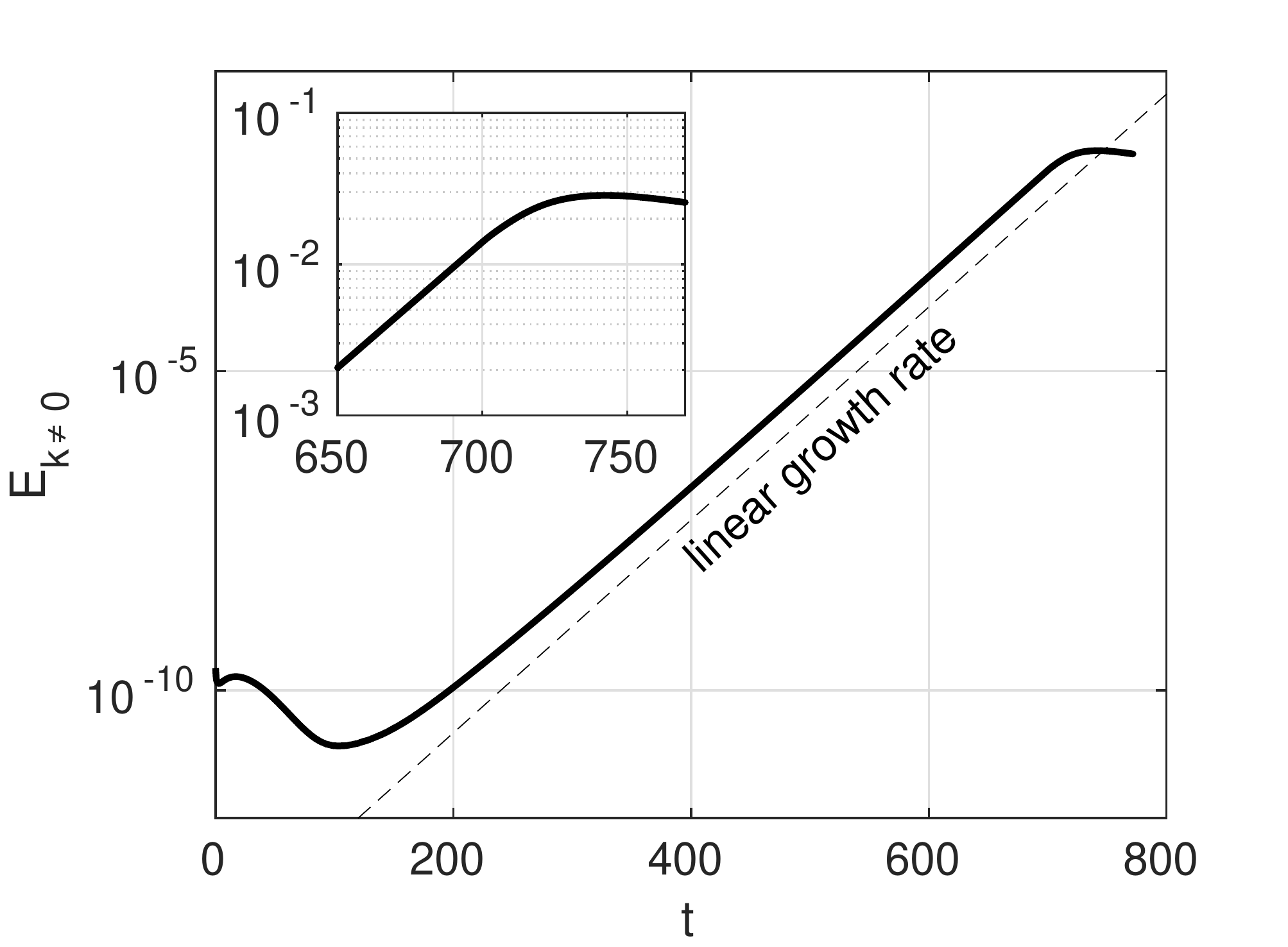}
%\vspace*{8pt}
\caption{The kinetic energy of the streamwise-dependent part of the perturbation velocity field, $E_{k\neq 0}$, starting from a base flow averaged in region I (see Fig. \ref{fig:domain_decomposition_mean_profiles}(a)). The initial condition is a noisy field of small amplitude ($\mathcal{O}(10^{-5})$). The inset shows a close-up of the transition from linear to nonlinear regime. \label{fig:energy_growth}}
\end{figure}

\begin{figure}
\centering
\includegraphics[width=0.8\textwidth]{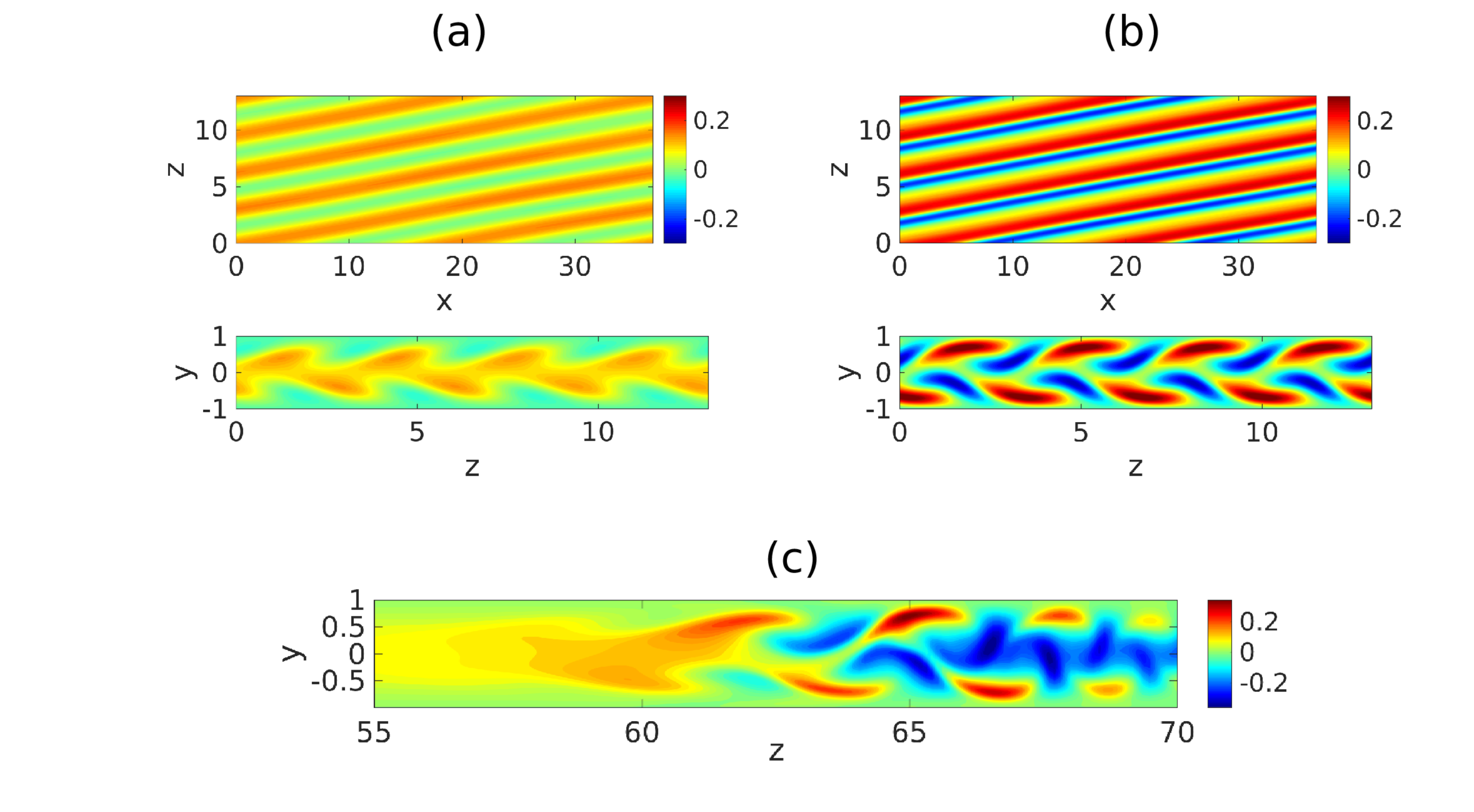}
%\vspace*{8pt}
\caption{The visualisation of the flow field of the simulation shown in Fig. \ref{fig:energy_growth}. (a, b) Flow fields at time-instant $t=650$ and 750 are plotted. The streamwise velocity \BS{with respect to the parabola $1-y^2$} in the $x$-$z$ cut-plane at $y=-0.5$ is plotted in the top panels and in the $z$-$y$ cut-plane at $x=0$ in the bottom panels. (c) The streamwise velocity perturbation with respect to \BS{the parabola} of the band shown in Fig \ref{fig:domain_decomposition_mean_profiles}(a) in the $z$-$y$ cut-plane at $x=57$, which cuts through the head of the band. \label{fig:nonlinear_development}}
\end{figure}

Figure \ref{fig:energy_growth} shows the kinetic energy of the streamwise-dependent part of the velocity field, $E_{k\neq 0}$. After some initial transients, the most unstable perturbation becomes dominant after $t=200$, undergoing a growth rate exactly given by our linear stability analysis (marked by the dashed line). The kinetic energy reaches an amplification of nearly $10^8$ from $t=200$ to 700. Later, nonlinearity kicks in and the kinetic energy starts to saturate after $t=700$. Figure \ref{fig:nonlinear_development} visualises the flow field near the end of the linear regime at $t=650$ (panel (a)) and at early nonlinear stage at $t=750$ (panel (b)). In comparison, the streamwise velocity of the real turbulent band in the $z$-$y$ cross-section at $x=57$ (see Fig. \ref{fig:domain_decomposition_mean_profiles}(a)), which cuts through the head of the band, is visualised in panel (c). From the comparison, we can see close resemblances between the flow structure in panel (a-b) and between $z=60$ and 67 in panel (c): both the pattern of the streaks in the $z$-$y$ plane and the spanwise wave length (or wave number) are very close. 

The linear analysis in Sec. \ref{sec:linear_analysis} and the nonlinear simulations indeed evidence the important role that the spanwise inflectional mean flow and the associated instability play in the streak generation at the band head.
\section{Discussion and conclusion}

Our simulations in large computational domains show that the head of a turbulent band travels into the adjacent laminar flow at a speed of 0.1 and the tail travels in the same direction as the head, therefore into the band body, at a speed relatively smaller than the head, which is roughly 0.06 in our simulation. Therefore, the band, as a whole, exhibits a slow drift in the spanwise direction at a speed of about $(0.1+0.06)/2$=0.08, which is close to the measurement of \citet{Tao2018,Shimizu2018}. Visualisation in Fig. \ref{fig:tail_large_domain} shows that the tail continually decays as a bulk, i.e., a substantially long patch containing multiple streaky structures decays as a whole. \BS{Our study also shows that, for DNS, a sufficiently large computational domain and long observation time are necessary to obtain reliable characteristics of a band, such as the speed of the tail and the tilt angle of the band.}

A close observation of the head indicates that wave-like streaky structures are continually generated at the head, see Fig. \ref{fig:streak_generation_first_streak}. We can conclude that the head is responsible, at least partially, for the growth and self-sustaining of the turbulent band. This is consistent with  \citet{Shimizu2018,Kanazawa2018} that a turbulent band is driven by an active head in this Reynolds number range. Considering the importance of the head, we particularly studied the streak generation mechanism at the head. Our results show that the (temporally and spatially averaged) mean flow at the head is strongly inflectional in the spanwise velocity component. We took the mean velocity profile at the head and carried out linear stability analysis, and indeed found an inflectional instability. The most unstable eigenvector gives a flow field that exhibits remarkable similarities with the streaky flow structure at the head, in terms of wave-like streaky flow pattern, spanwise wave length (or wave number), wave speed and the tilting direction with respect to the streamwise direction. A further DNS study of the linear and nonlinear development of small perturbations also shows close resemblances with the flow pattern at the head of a real turbulent band (see Fig. \ref{fig:nonlinear_development}).  
Both the linear stability analysis and nonlinear simulation strongly suggest that the mechanism of the streak generation at the head is the spanwise inflectional instability associated with the local mean flow. 

\BS{\citet{Henningson1987,Henningson1989} proposed a very similar `growth by destabilisation' mechanism for the growth of turbulent spots at higher Reynolds numbers. \citet{Dauchot1995} also proposed similar mechanism for the growth of turbulent spots in plane Couette flow. In addition, \citet{Hof2010} proposed that an inflectional instability at the upstream edge of turbulent puffs is responsible for the turbulence production and therefore the self-sustainence of puffs in pipe flow, although \citet{Shimizu2009} proposed a different mechanism via Kelvin-Helmholtz instability. Our study here provides more quantitative evidences for the inflectional instability being the underlying mechanism in the case of turbulent bands at low Reynolds numbers. All these studies suggest that this inflectional instability associated with the local mean flow is a rather general mechanism for the growth of localised turbulence in wall-bounded shear flows.}

\BS{It should be noted that, although our linear analysis captures some key characteristics of the head, our analysis does not quantitatively capture some other characteristics
of the waves at the head. For example, the tilting angle of the waves is about 38 degrees,
whereas the most unstable mode of our linear instability analysis only shows an
tilt angle of about 10 degrees. 
Besides, we see a relatively fast growth of streaks (see Online Movie 1), while the modal growth obtained in our analysis seems to be slow. 
These deficiencies probably can be attributed to the over-simplification of
the linearisation about a temporally and spatially averaged mean flow, during
which many aspects of the real flow have been dropped. For example, the counter-clockwise large scale flow around the head (see the vectors in Fig. \ref{fig:domain_decomposition_mean_profiles}(a)) could potentially contribute to the large tilt angle of the streaks at the head. Clearly this effect cannot be accounted for in our simplified analysis. A more rigorous stability analysis taking into account the temporal and spatial dependence of the mean flow is needed to more quantitatively capture all these characterisitics.}

\BS{Based on our analysis, we would like to propose} a self-sustaining mechanism of \BS{fully localised} turbulent bands at low Reynolds numbers, \BS{which is similar to the self-sustaining mechanism of the wingtips of turbulent spots proposed by \citet{Henningson1989}}: Spanwise inflectional instability generates streaks, which are quickly amplified and become turbulent and move away from the head (towards the tail, see Online Movie 1). The streak generation rate is higher than the decaying rate of streaks at the tail, therefore, the band can achieve transverse growth. In turn, the band maintains a large scale flow around the band (via a similar mechanism as described by \citet{Duguet2013}, which bears a spanwise inflectional instability at the head of the band. \BS{A noticeable difference with the spot case is that streaks generated at the head largely keep a form of traveling wave with relatively weak turbulent fluctuations while traveling within the bulk of the band (see the flow pattern in Fig. \ref{fig:band_and_head_tail_speed}(a), Fig. \ref{fig:tail_large_domain} and the Online Movie 1).}

\BS{However, it has been shown that turbulent bands can be sustained either in a small tilted domain by \citet{Tuckerman2014} or even in large domains by \citet{Tao2018} if the band is periodic and no head and tail exist. Further study is needed to elucidate how streaks in the bulk of the band sustain themselves and maintain a traveling wave-like form with turbulent fluctuations on top. For plane Couette flow, \citet{Rolland2015, Rolland2016} investigated the sustaining mechanism of the turbulent streaks within turbulent bands and spots and concluded that the shear layers inside the velocity streaks generate vorticity and are responsible for the self-sustainence of the turbulent streaks. \citet{Duguet2013} described the motion of the streaks inside turbulent bands in plane Couette flow as the advection of small-scale structures by the large-scale flow. It is interesting to investigate the case in channel flow following their analysis. Another open question regards what mechanism determines the tilt angle of the band. We speculate that the wave speed of the waves generated at the head, with respect to the convection speed of the bulk of the band, may give a hint to the answer.}

\section{Acknowledgements}
We acknowledge the financial support from the National Natural Science Foundation of China under grant number 91852105 and from Tianjin University under grant number 2018XZ-0006. We acknowledge the computing resources from National Supercomputer Center in Guangzhou. Helpful discussions with Jianjun Tao, Dwight Barkley and Ashley P. Willis are highly appreciated. 

\bibliographystyle{jfm}
%\bibliography{references}

\end{document}